\documentclass[10pt]{article}
\usepackage{amsmath}
\usepackage{amssymb}
\usepackage{graphicx}
\usepackage{algorithm2e}

\usepackage{cite}

\usepackage{color} 

\usepackage[labelfont=bf,labelsep=period,justification=raggedright]{caption}

\makeatletter
\renewcommand{\@biblabel}[1]{\quad#1.}
\makeatother

\date{}

\begin{document}

% Title must be 150 characters or less
\begin{flushleft}
{\Large
\textbf{Neuronal functional connectivity among multiple areas of the rat somatosensory system during spontaneous and evoked activities}
}
% Insert Author names, affiliations and corresponding author email.
\\
Antonio G. Zippo$^{1}$, 
Riccardo Storchi$^{2}$, 
Giuliana Gelsomino$^{1}$,
Sara Nencini$^{1}$,
Gian Carlo Caramenti$^{3}$,
Maurizio Valente$^{1}$,
Gabriele Eliseo M. Biella$^{1,\ast}$,
\\
\bf{1} Institute of Molecular Bioimaging and Physiology, National Research Council, Segrate, Milan, Italy
\\
\bf{2} Faculty of Life Science, University of Manchester, Manchester, UK
\\
\bf{3} Institute of Biomedical Technology, National Research Council, Segrate, Milan, Italy
\\
$\ast$ gabriele.biella@ibfm.cnr.it
\end{flushleft}

% Please keep the abstract between 250 and 300 words
\section*{Abstract}
Small-World Networks (SWNs) represent a fundamental model for the comprehension of many complex man-made and biological networks. In the central nervous system, SWN models have been shown to fit well both anatomical and functional maps at the macroscopic level.  However the functional microscopic level, where the nodes of a network are composed of  single neurons, is still poorly understood. At this level, although recent evidences suggest that functional connectivity maps exhibit small-world organization, it is not known whether and how these maps, potentially distributed in multiple brain regions, change across different conditions, such as spontaneous and stimulus-evoked activities.\\
We addressed these questions by simultaneous multi-array extracellular recordings in three brain regions diversely involved in somatosensory information processing: the ventropostero-lateral thalamic nuclei (VPL), the primary somatosensory cortex (S1) and the centro-median thalamic nuclei (CM). From both spike and Local Field Potential (LFP) recordings, we estimated the functional connectivity maps by using the Normalized Compression Similarity (spikes) and the Phase Synchrony (LFPs). Then, by using graph-theoretical statistics (clustering coeffient, characteristic path length, small-worldness), we characterized the functional map topology both during spontaneous activity and sensory stimulation.\\
Our main results show that: (i) spikes and LFPs show SWN organization during spontaneous activity; (ii) After stimulation onset, while substantial functional map reconfigurations occur both in spike and LFPs, small-worldness is nonetheless preserved (iii) The stimulus triggers a significant increase of inter-area LFP connections without modifying the topology of intra-area functional connections; (iv) Through computer simulations of the fundamental concept of cell assemblies, transient groups of activating neurons can be described by small-world networks.\\ 
Our results suggest that neural activity among neurons from multiple areas of the rat somatosensory system allows for the integration of local computations occurred in distributed functional cell assemblies according to the principles of SWNs.

\section*{Author Summary} 
Cell assemblies (sequences of neuronal activations), seem to represent a functional unit of information processing. However, it remains unclear how groups of neurons may organize their activity during information processing, working as a sole functional unit. One prominent principle in complex network theory is covered by small-world networks, in which nodes are easily reachable by each other and are organized in highly dense clusters. Small-world networks are already found on large-scales in human and primate brain areas while their presence at the neuronal level remains unclear. The aim of this work was to  investigate the possibility that functional, related neural populations, encompassing multiple brain regions, can be organized in a small-world network. We investigated the coherent neuronal activity among multiple rat brain regions involved in somatosensory information processing. We found that the recorded neuronal populations represented small-world networks maintained during stimulations. Furthermore, by using computer simulations, we inferred that small-world networks represent a plausible topology for cell assemblies.\\
This work suggests that the coherent activity of neurons from multiple brain areas in somatosensory system, which can comprise individual functional units, promotes the integration of local computations, the functional principles of small-world networks.

\section*{Introduction}
Neurons in the brain form highly composite networks sustained by a complex and variously distributed thread of synapses. Although the detailed anatomical connections are still under investigation \cite{sporns} it has been shown that, during in-vivo recordings, active neurons form functional assemblies that do not necessarily depend on the underlying anatomical connectivity \cite{feldt}. Therefore, while anatomical connectivity is stable for relatively long times, the functional connectivity is highly dynamic and depends on the particular tasks triggered by internal and external events. Critically, the organization principles that control the functional connectivity among single neurons and small neuronal populations are still poorly understood \cite{bassett3}. The early hypotheses \cite{hebb}, that distributed groups of neurons can operate as a functional unit through coordinated activities, have now been supported in observations of transient cell assemblies over different cortical regions \cite{canolty}. It remains, however, unclear how these dispersed neurons may gather in a functional unit for information processing.\\
In the last 15 years, thanks to a substantial advancement in the complex network theory, Small-World Networks (SWNs) emerged as a paradigmatic model and provided the analytical tools to explain a large set of complex networks in the most diverse scientific areas \cite{watts,barabasi}. Furthermore, it has been suggested that small-world networks represent optimized structures accomplishing adequate balance between information transfer efficiency and reliability \cite{bassett1,bassett2,barabasi,vertes,barrat,barahona}. Following this trend, large networks of brain areas, when studied from a macroscopic perspective by imaging techniques, have been shown to functionally organize as SWNs \cite{horwitz,bullmore}. Contrarily, the functional connectivity emerging at the microscopic level of single neuron networks is still poorly understood, although recent works provided evidence of SWNs in small local neuronal populations \cite{bettencourt,humphries2006,yu,humphries2008,pajevic,varshney,takahashi,gerhard,downes}. Specifically it is not clear whether functional groups of neurons from multiple brain regions display a SWN organization.\\
In the present work we address these points by means of spike and LFP simultaneous multi-array recordings from the ventropostero-lateral thalamic nucleus (VPL), the centro-median thalamic nucleus (CM)  and, the primary somatosensory cortex (S1) \cite{vanDerWerf,sherman2,grossberg}. We analyzed VPL, CM and S1 because these regions are actively involved in the tactile information processing in somatosensory system of mammal brains. Because spikes and LFPs represent a dualistic picture of what (respectively) neurons and local neuronal populations are doing, we decided to test the working hypothesis for both signals. To estimate the strength of functional connectivity in spiking activity we introduced a novel function that allows us to detect long-range temporal dependencies which occur in thalamocortical interactions (NCS) that are otherwise unrecognizable by correlation analyses. Instead, the LFP functional connectivity was estimated by a standard measure of phase synchrony.\\
Our results confirm the presence of SWNs in crucial stations of the rat somatosensory system, during spontaneous activity and provide evidence that, during stimuli, small-world organization principles are preserved in spite of massive topological reconfigurations. In addition, further results obtained by computer simulations showed that the observed functional conditions  may be the expression of a large-scale functional substrate composed by small neural groups with small-world topologies where node membership to each group is dynamical.\\
Consistent with other findings on large-scale brain networks \cite{canolty,siegel,womelsdorf,fries}, our results provide evidence for a model of functional organization where distinct functional neuronal assemblies are sparse and encompass multiple brain areas.

\section*{Results}
\subsection*{Functional Connectivity in Spiking Activity}
Our first aim was to estimate the topology of functional connectivity graphs obtained from simultaneous spiking activities of neuronal populations in VPL, CM and S1. Because specific tactile stimulation can potentially exert a powerful effect on functional connectivity, we set out to evaluate functional connectivity graphs both during spontaneous and evoked activities. In order to compute the graph's connectivity in the two conditions, we performed pairs of 10 minutes recordings, the first with no stimuli, the second by delivering fast tactile pulses (see methods) at randomized intervals (500 $\pm$ 200 ms) on the five digits.\\
We estimated the functional connectivity in neuronal spiking activity by using the Normalized Compression Similarity ($NCS$) function to detect functional couplings between pair of neurons. $NCS$ is defined in the range $[0,1]$, where $0$ indicates no interaction and $1$ indicates an exact correspondence between the firing patterns of the two neurons considered. This measure was chosen in place of more conventional ones because of its ability to capture both short-range (synchronous) and long-range (delayed) interactions between neurons. Its efficacy was then assessed by analyses on synthetic spike train (see Methods, Fig. \ref{fig1}C-E). We thus scanned the recorded activity in search for functional relations by using sliding windows of different lengths (50, 250, 500, 1000 ms). The time length of a sliding window defined the largest possible delay at which an interaction could be detected (see Methods). After the $NCS$ estimation on all possible pairs of simultaneously recorded neurons, the resulting adjacency matrices were binarized by a threshold in order to obtain the functional connectivity graphs. Subsequently, the small-world statistics $C$ (clustering coefficient), $L$ (characteristic path length) \cite{watts}, $S$ and $\omega$ were estimated on these graphs \cite{humphries2008,telesford}. The first measures the tendency of neurons to segregate into separate sub-networks, the second measures the average path length between nodes, the third and the fourth are two measures of small-worldness. The terms $C^r$ and $L^r$ represent normalizing values estimated by algorithms which randomize the original networks but preserving the node degree distributions. Finally, we further characterized the resulting graphs by analyzing the node degree distribution, the centrality and the community structure. Specifically, we wondered if stimuli provoked changes in the node degree or the node centrality distribution or if they affected the number of communities.\\
First, we measured the small-worldness statistics in our spontaneous activity recordings and we found that S and omega stated that such graphs can be considered small-world networks ($S > 1$, $\omega$ close to $0$), irrespective of the time window used for detecting the functional connections (50, 250, 500, 1000 ms; Table \ref{table1}).\\
To note that increasingly larger windows make smaller $NCS$ values because statistical dependencies become more and more sporadic. Consequently, the threshold decreased when windows became larger. Time windows larger than 1 second reported no small-world network configurations.
Then we compared the average functional connectivity in time windows of 100 ms duration before and after the stimulus onset. 100 ms was a reasonable time constraint in order to consider most of the thalamocortical interactions. The functional connectivity graphs underwent significant topological reconfigurations after the tactile stimulus delivery. As an example, in Fig. \ref{fig2}, the neuron 32 switches from a fully disconnected state (Fig. \ref{fig2}A, relative to pre-stimulus condition) to highly connected (hub) state (Fig. \ref{fig2}B) after the stimulus onset. Surprisingly, notwithstanding these profound changes, the small-worldness computed on the pre- and post-stimulus functional connectivity graphs were not significantly different ($P > 0.13$ in all comparisons, non-parametric Wilcoxon ranksum test, Table \ref{table2}). Therefore, the average small-world properties were not perturbed by tactile stimulations. 
Furthermore, we performed a correlation analysis between the quantitative changes in functional connectivity graphs and the firing rate variations induced by tactile stimuli. We found that the changes induced by stimuli were significantly greater ($P < 0.005$, ranksum test) than those observed in spontaneous activity. This means that stimuli effectively modulated the functional connections among neurons. In fact, the firing rate of neurons were positively correlated with the evoked connectivity changes ($R=0.67$, $P < 0.001$, t-test; Fig. \ref{fig3}A). Namely, whenever the recorded neurons were effectively involved in the response to stimuli, these stimuli recruited concurrent functional connection changes proportional to the evoked firing rates.\\
 Subsequently, we analyzed the extracted graphs in pre- and post-stimulus conditions by their node degree distributions that were not significantly different (Figure \ref{fig3}B, $P=0.48$, ranksum test). Moreover, by analyzing the distribution of node centrality, we found that graphs in pre-stimulus conditions (mean, $\mu = 22.84$, standard deviation, $\sigma=1.63$) had smaller betweenness centrality ($P<0.000$, ranksum test) than graphs in post-stimulus conditions ($\mu=25.67$, $\sigma=2.01$) indicating that stimuli made more centralized nodes. Moreover, by further analyses we found that the betweenness centrality was not equally distributed over the three regions (VPL,S1,CM) both in pre- and in post-stimulus conditions: performing an ANOVA-1-way test over the distributions of centrality in the three regions, we found that these had not the same mean ($P = 0.007$ in pre-stimulus, $P = 0.002$ in post-stimulus). Furthermore, the Figure \ref{fig3}C shows that the centrality of neurons from CM increased ($P=0.013$, ranksum test), decreased for VPL neurons ($P = 0.000$, ranksum test) and increase for S1 neurons ($P=0.006$, ranksum test). This suggested that neurons from S1 received more load in the processing of stimulus information.\\  
By counting the number of communities, we found that graphs in pre-stimulus conditions ($\mu=5.04$, $\sigma=1.45$) had less communities ($P<0.01$, ranksum test) than graphs in post-stimulus conditions ($\mu=3.42$, $\sigma=1.13$). Hence the incoming stimulus information forces the functional networks to reduce the number of communities indicating that more neurons were involved in the stimulus representation.

\subsection*{Functional Connectivity in LFPs}
Our second aim was to estimate the topology of functional connectivity graphs obtained from the LFP activity recorded simultaneously to the spiking activity. We estimated the LFP functional connectivity maps by following the same sequence of analyses used for spiking activity. First, we estimated the functional connectivity between all possible electrode pairs in order to generate the adjacency matrix, then we binarized this matrix by using a variable threshold (see Methods) and finally we computed the network statistics.\\
In order to estimate the functional connectivity between pairs of channels we used the phase synchrony measure $\gamma$ (see Methods) \cite{montemurro}. Phase synchrony is more suitable than NCS for continuous signals and it has been widely applied to EEG and LFP analyses \cite{pereda}. It is normalized in the range $[0,1]$, where unity occurs when phase coupling is exact and zero when it is null.\\
For spontaneous activity recordings, LFP functional connectivity graphs exhibited time-invariant small-world properties (Table \ref{table3}). 
Even in this analysis, increasingly larger windows makes smaller $\gamma$ values thus the threshold  were decreased when windows became larger. Time windows larger than 1 second reported no small-world network configurations. We wondered if the synchrony was crucial for LFP phases or small-worldness can emerge even without tight time constraints. By using NCS rather of gamma function to estimate functional connectivity, small-world networks did not appeared (data not shown).\\
For evoked activity recordings, stimulus occurrence triggered significant  topological reconfigurations in functional graphs, as shown, as a representative case, in Fig. \ref{fig4}. The picture reports a functional connectivity graph estimated before (Fig. \ref{fig4}A) and after (Fig. \ref{fig4}B) the stimulus onset. During spontaneous activity, CM (green), VPL (blue) and S1 (red) showed tight intra-area synchronization and poor inter-area synchronization. After the stimulus onset inter-area synchronization emerged while intra-area synchronization was maintained roughly constant.\\
We asked whether these observations on a single recording could generalize to our full dataset. In order to test for this possibility we first divided our LFP recordings into two classes: responsive and non-responsive. LFP recordings were considered responsive when the average evoked firing rate, measured from the same recording channels (see Methods), was larger than the mean plus 5 times the standard deviation of the basal firing rate. We found that the number of intra-area (local) functional connections was preserved across the three conditions of spontaneous, non-responsive and responsive LFP activity ($P = 0.63$ for CM, VPL and S1, ranksum test). However, during responsive LFP recordings, the average number of inter-area (or global) functional connections was substantially larger for all possible inter-area combinations ($P < 0.002$ for CM-VPL, CM-S1, VPL-S1, ranksum test).\\
Subsequently, we compared pre- and post-stimulus functional graphs by computing the networks statistics in 50 and 100 ms windows before and after the onset of tactile stimuli. We challenged both windows size to establish the possible dependency of diverse time constraints during synchronization event. We found that both conditions (pre- and post-stimulus) exhibited small-world properties and the statistics were not different (Table \ref{table4}, $P > 0.38$ in all comparisons, ranksum test) and the window size per se did not change the small-worldness values. Interestingly, the tactile stimulus onset triggers a substantial increase in the number of inter-area connection but does not have a significant effect on the intra-area connections (Fig. \ref{fig5}A). We concluded that small-world statistics are preserved in LFPs during both spontaneous and evoked activity. Thereafter, we analyzed the extracted graphs in pre- and post-stimulus conditions by their node degree distributions that were not significantly different (Figure \ref{fig5}B, $P=0.83$, ranksum test).\\
By analyzing the distribution of node centrality, we found that graphs in pre-stimulus conditions ($\mu=21.46$, $\sigma=2.17$) had smaller betweenness centrality ($P <0.03$, ranksum test) than graphs in post-stimulus conditions ($\mu=23.79$, $\sigma=1.63$) indicating that stimuli cause nodes to be more recruited. Furthermore, we found that the betweenness centrality was not equally distributed over the recorded regions (VLP, S1, CM) both in pre- and in post-stimulus conditions: performing an ANOVA-1-way test on the distributions of centrality in the three regions, we found that they had not the same mean ($P = 0.032$ in pre-stimulus, $P = 0.039$ in post-stimulus). Furthermore, the Figure \ref{fig5}C shows that the centrality in CM decreased ($P=0.015$, ranksum test), increased for VPL populations ($P = 0.033$, ranksum test) and decrease for S1 neurons ($P=0.013$, ranksum test). This suggested that neural populations in VPL received a more network load in stimulus processing.\\ 
Moreover, by counting the number of communities, we found that graphs in pre-stimulus conditions ($\mu=2.97$, $\sigma=0.71$) had less communities ($P<0.004$, ranksum test) than graphs in post-stimulus conditions ($\mu=2.14$, $\sigma=0.67$). In conclusion, the incoming stimulus forces delegated networks to merge the active communities indicating that more neural groups expressed coordinated oscillations.

\subsection*{Functional Connectivity in Simulated Spiking Activity}
In the previous sections we provided evidence, both at a pre- and post-synaptic level, that cortical and subcortical neurons can be functionally organized as small-world networks. Our results are consistent with recent findings from large scale cortical recordings \cite{pawela,bassett1,zhang} and suggest the existence of sparse functional networks composed of neuron assemblies that encompass multiple cortical and subcortical areas; importantly, each functional network would recruit its units not necessarily in close proximity \cite{fries,canolty}. However, it remains unclear what kind of functional substrate supports the observed networks, i.e. how large-scale network of neurons are organized such that sampling from a small subset of nodes, we can observe small-world configurations.\\
To address this question, we investigated the hypothesis that neuronal networks are arranged in functional groups, recalling the concept of cell assembly, where the membership to each assembly is dynamical \cite{hebb} and assembly neurons can be spatially distributed. To evaluate this conjecture we developed an artificial neuronal network where nodes are arranged in dynamical groups and emit spikes following small-world criteria. Subsequently, we sampled the activity of small ensembles of nodes and compare the network statistics from the synthetic spiking activity with those obtained by recordings. Once asserted such hypothesis, we estimated how many functional groups can coexist within a neural network keeping consistency with experimental observations.\\
For these purposes, we implemented a large-scale simulation of $N$ neurons that embeds $k$ small-world networks. Then we estimated small-worldness by systematically varying $k$. We simulated a neural network of $N = 100000$ neurons, about five fold the estimated size for a rat cortical column in the somatosensory system \cite{meyer}. The size of each functional group was chosen by sampling from a discrete uniform distribution $\sim U(50, N/100)$. This distribution and its parameters are consistent with the results of \cite{mountcastle}. For further details about the implementation see the relative Methods section.\\
At each run we randomly extracted 100 units (comparable with the number of units that we could simultaneously record during the experiments) and we computed the network statistics. We observed that small-world statistics underwent a bimodal behavior (Table \ref{table5}), increasing for $k = 0.01$, peaking when $k = 0.1$ and decreasing with $k$ up to $0.5$. At peak we found comparable statistics values with those measured experimentally both on spiking ($P = 0.60$, ranksum test) and LFP activities ($P = 0.46$, ranksum test). From our results we conclude that neural activity may be structured in overlapping functional groups, each of them organized as a small-world network. Finally, we extracted admissible values (which keeps consistency with $S$ and $\omega$) of small-world networks per number of nodes that ranged within $0.01$ and $0.1$.

\section*{Discussion}
In this paper we show that spiking and LFP activities of neurons, in three stations of the somatosensory system of rat brains, present clear signs of small-world functional organization with sub-second invariance. Furthermore, we show that this distinctive functional organization persists in the presence of tactile stimuli, independently of the neural response intensity. Finally, results obtained by means of computational models suggest that small-world networks may represent a consistent and formal model for cell assemblies.

\subsection*{Small-world in brain networks}
Studies in anatomical brain connectivity discovered that small-world network architectures are a distinctive trait in animals \cite{yu,perin,varshney}, yet with different degree of brain development, including humans \cite{bassett1,honey,bullmore}. The complementary observation that brain pathologies like epilepsy \cite{liao} and schizophrenia \cite{micheloyannis} show small-world network disruptions provides further support to the potential interpretive and explanatory strength of small-word topology. 
It should be noted that the methodological approaches used to study the functional connectivity at macroscopic and microscopic levels are slightly different. Studies with fMRI estimate functional connectivity comparing the BOLD signal of each node to a seed voxel whereas the functional strength between neurons is computed comparing the electrical activity of all possible neuron couples \cite{zalesky}. Setting apart the technical incongruences and focusing on neuronal functional connectivity and its underlying topology, a number of beautiful studies found an intrinsic small-world topology in single visual areas of primates \cite{bettencourt,humphries2006,pajevic,takahashi,yu,gerhard} and in neuronal cultures \cite{downes}. These elegant approaches gave access to the richness of the local functional connectivity scaled at the neuronal level. However, none of these studies provided a description of an extended connectivity, involving different neuronal populations nested in close or far brain regions.\\ 
Yet, according to the work of Bassett et al. \cite{bassett2}, it is suggested that the functional organization in small-word networks may be ubiquitous at different spatial and temporal scales and different experimental conditions. This organization could represent a stable, and even a necessary, scheme for information processing in brain areas, networks and neurons.

\subsection*{Tactile information processing}
In this study our aim was to explore the functional connectivity of different brain areas involved in somato-sensory information processing and to envisage a potential broadening to other brain circuits. As it is known, the precise mechanisms regulating the rich and complex neural thread of brain areas involved in the construct of tactile processing are still far from being clarified. However, a response to sensory inputs is well recognized for a number of areas, well including the areas we chose for this study. Namely, several studies showed that mechanoreceptor signals reach both VPL and CM thalamus. Then, the former outputs directly to innervate S1, while the latter outputs are more diffused and addressed to the higher sensorimotor cortical layers \cite{sherman1,sherman2,grossberg,swadlow,sadikot}. We observed that in the resting state, VPL, CM and S1 showed substantial mutual connectivity and, in addition that, under stimuli, they establish an even more integrated architecture enabling fast information exchanges. This functional connectivity could be ascribed to the necessity to create larger functional units and nodes with higher network load. In accompaniment of these changes, however, the topology of the small-world networks never degenerated. This dualistic behavior, redistribution of connectivity and permanence of topology, appears an interesting interpretive key, potentially extendable to other complex brain networks and useful to analyze also pathological signs in brain dynamics, like epilepsies or schizophrenia, where the detection of disrupted network or loss of topological hallmarks may help novel nosological classifications.

\subsection*{Cell assemblies and the synthetic modell}
The analyses from computational models suggest that a cell assembly, i.e. a transient functional unit composed by neurons potentially distributed in separated brain regions \cite{hebb}, appear to be organized as a small-world network. The null hypothesis that such groups were random networks does not match experimental evidence.\\    
Our model assumed that neurons may variably join in concurrent assemblies and we empirically estimated that the admissible number of assemblies that a neuron could participate in, covered a range between 0.01 and 0.1. It could be objected that the number of shared neurons appears to be relatively high. However, it might be noted that neurons in stimulated conditions probably get a balance between the number of assemblies and the inherent metabolic cost \cite{fonseca-azevedo}. As a consequence, neuron sharing could abate the metabolic cost of network activations, behaving as a parsimonious limiting factor to an unregulated growth of recruited functional units.\\ 
Our synthetic model tried to answer a double hypothesis: the first that the topology expressed by our small graphs could be a nested and natural expression of a homologous larger population. The reverse question appears to be if the hypothetical extended topology that we designed in our synthetic model could suitably host a subset of units with the described topological properties of our recorded networks. Namely, we argue that there are mutually fitting topological features between our recorded and the hypothetical larger synthetic networks.\\
Our results are also strictly related research suggesting that neuronal oscillations enable selective and dynamic control of distributed functional cell assemblies \cite{canolty}. We could enrich such a scenario, speculating that LFP coherent activities may reflect the integration of local computations which occurred in these distributed cell assemblies. Thus, the small-world topology expressed by synchronized and distributed LFP phases would support a hierarchical and efficient functional substrate for incorporating cell assemblies.

\subsection*{Limitations and developments}
The use of gaseous anesthetics represents a potential limitation of this study. The level of Isoflurane was indeed very low and, as it comes from other studies, low enough to avoid important suppressive actions of the neural activity \cite{wang}. However, it is implicit that conclusive studies with waking animals are needed to definitely address the existence of these functional topologies (at least) in these three brain regions.\\
We proved the effectiveness of the NCS measure to capture the long-range dependencies and although such method is inspired by related approaches \cite{london,szczepanski,amigo,zippo}, it has never been applied before on electrophysiological data. A potential drawback of such approach is represented by its computational cost. Indeed, a numerical increase of recorded neurons (up to thousands or more), would need a cubic increase of the computational time required to calculate the NCS adjacency matrices.\\
Ultimately, inferences from computational models are often risky. In fact, our findings mainly endorse but not prove the hypotheses. Other topologies, rather than random networks, could deliver consistent results. A set of other debated topologies (hierarchical modular networks, scale-free networks, etc.) could be also considered.

\subsection*{Conclusive notes}
In an evolutionary perspective, small-world topologies appear preferentially selected among network topologies under the natural constraints (efficiency and reliability) of brain expanding complexity in the mammal phylogenetic tree \cite{vertes}. It can satisfy the necessity of internal input integration and grants for best responses to environmental requirements. Moreover, small-world networks seem coherent with the recent advancements in the physiology of neuronal networks. More punctually, from a functional point of view, small-world networks appear to provide dynamical features, such as communication-through-coherence \cite{fries,womelsdorf,canolty}, for fast information integration where even far located nodes efficiently participate to the information process.

\section*{Materials and Methods}

\subsection*{Ethical notes}
All the animals have been used in accordance to the Italian and European Laws on animal treatment in Scientific Research (Italian Bioethical Committee, Law Decree on the Treatment of Animals in Research, 27 Jan 1992, No. 116). The National Research Council, where the experiments have been performed, adheres to the International Committee on Laboratory Animal Science (ICLAS) on behalf of the United Nations Educational, Scientific and Cultural Organizations (UNESCO), the Council for International Organizations of Medical Sciences (CIOMS) and the International Union of Biological Sciences (IUBS). As such, no protocol-specific approval was required. The approval of the Ministry of Health is classified as ``Biella 1, 3/2011'' into the files of the Ethical Committee of the University of Milan.

\subsection*{Animal preparation and stereotaxis}
Seven male albino rats (Sprague-Dawley, Charles River, Calco, LC, Italy, $300-400$ g) were chosen in the set of 11 animals employed in the recordings. All the animals were maintained in a $16/8$ hour light-dark cycle with access to food and water ad libitum. The rats underwent preliminary barbiturate anesthesia for the surgical experimental preparation. The jugular vein and the trachea were cannulated to gain direct drug delivery access and a connection to the anesthesia-ventilation device. Before the placement of electrodes, rats were paralyzed by intravenous Gallamine thriethiodide ($20$ mg/kg/h) injection and connected to the respiratory device delivering (1 stroke/s) an Isoflurane ($2.5\% 0.4$ to $0.8$ l/min) and Oxygen ($0.15-0.2$ l/min) gaseous mixture \cite{kohn}. Curarization was maintained stable throughout the whole experiment by Gallamine refracted injections ($0.1$ ml of the original solution/h). 
The anesthesia levels were maintained into ranges which prevented any corneal or retraction reflex (in absence of curarization) with low intensity noxious mechanical stimuli applied on a posterior paw \cite{kohn}.\\
We chose three areas for the simultaneous neuronal recordings in the left brain: the thalamic median nuclei (in particular the centromedial (CM)), the thalamic ventro-postero-lateral nuclei (VPL) \cite{sherman1,sherman2} and the primary somatosensory (S1) cortex. Fast tactile stimuli were delivered to the right posterior paw (see Fig. \ref{fig1}F) by a suitable electromechanical device.\\
Two holes were drilled on the skull. A $3\times2$ mm bone window for the access of the cortical matrix electrodes and a larger bone window ($6\times2$ mm) allowing for the simultaneous insertion of two parallel electrode matrices directed to the thalamic nuclei. The cortical access was set around a reference at $-1.5$ mm AP and $-2.5$ mm ML on the left, \cite{paxinos} and the electrode matrix was driven around $450$ to $800$ micrometer deep by an electronically controlled microstepper (Narashige, Japan). The thalamic access was centered at the two focus points of $-6$ mm AP, $-0.8$ and $-2.5$ mm ML \cite{paxinos}. The electrodes were inserted with a 25$^o$ slant and driven at least to $5500 \mu$m in depth and then advanced by a second electronically driven microstepper (AB Transvertex, Stockholm) until responses were observed to peripheral test stimuli. The neuronal recordings were obtained with two types of matrices, a vertical array devoted to the cortical recordings and two planar matrices devoted to the 
thalamic recordings. The vertical array was a multitrode (Multitrode Type 1, Thomas RECORDING GmbH, Giessen, Germany) with $8$ gold contacts ($125 \mu$m contact spacing) with an average impedance of $1.2$ M$\Omega$. For planar matrices were $3\times3$ frames of tungsten or Pt-Ir electrodes, inter-tip distance $150-200 \mu$m, tip impedance $0.5-1$ M$\Omega$ (FHC Inc., ME, USA).\\
Fast thalamic and cortical responses to light tactile stimuli in the plantar aspect of the right hind limb were used as anatomo-functional acceptance criterion for acquisition.

\subsection*{Tactile Stimulation}
Controlled stimulation was delivered through a blunted cactus thorn on each of five sites of the rat right hind limb (Fig. \ref{fig1}F). The tip was mounted on the dust cap of a speaker and driven through an Arduino microcontroller board (available at http://www.arduino.cc). At the beginning of each stimulation epoch the tip was lightly placed over the skin. Fast $5$ ms pressure pulses were applied following a semi-random sequence. Pulses occurred in couplets. The delay between the first pulses of each couplet was set at $500$ ms. Every second pulse of each couplet followed the first by a random delay extracted uniformly in the range $150-250$ ms (see Fig. \ref{fig6}C). The stimuli semi-randomness was adopted to avoid habituation \cite{storchi}.

\subsection*{Neurophysiological recordings and preliminary data analyses}
For signal amplification and data recordings a $40$ channel Cheetah Data Acquisition Hardware was used (Neuralynx, MT, USA, sampling frequency $32$ kHz). Electrophysiological signals were digitized and recorded with bandpasses at $6$ kHz/$300$ Hz for spikes, $180$ Hz/1 Hz for Local Field Potentials. The data stored were analyzed off-line both using Matlab and by locally developed software. The neural firing rates had a mean of $31.4$ Hz with standard deviation of $26.8$ Hz. 
After the recordings the LFPs were downsampled to $0.5$ KHz. We used for filtering the same techniques described in \cite{zanos}. After filtering and downsampling, the spike contamination of LFP signals was null avoiding further spike removal techniques \cite{montemurro}. The spikes were extracted and sorted by using the \texttt{Wave\_clus} MATLAB toolbox \cite{quiroga}. Sorted cells with average rates below $4$ Hz and above $100$ Hz were excluded from the analysis. Furthermore, neurons resulted from sorting which had improbable inter-spike-interval distributions were discarded as well. Recorded neurons were uniformly distributed over the recording matrices and every electrode show distinct neural activity otherwise the matrix was repositioned. At the end of this process, we collected a total of $391$ neurons ($56 \pm 17$ in each experiment) out from the set of the acquired signals. Distribution of firing rates and inter-spike intervals is shown in the Supplementary Figures \ref{fig6}A-B.\\
The timestamps of spike occurrences were represented by binary sequences where 1's labeled a spike. We considered time bins of 1 ms thus avoiding occurrence of multiple spikes within the same bin.  Finally, we split each sequence into fixed-length (from $50$ to $1000$ ms) overlapping windows (Fig. \ref{fig1}C), thus obtaining an ordered set of equal length windows.

\subsection*{Functional connectivity by spike-train similarities}
\begin{figure}[!h]
\includegraphics[width=\textwidth]{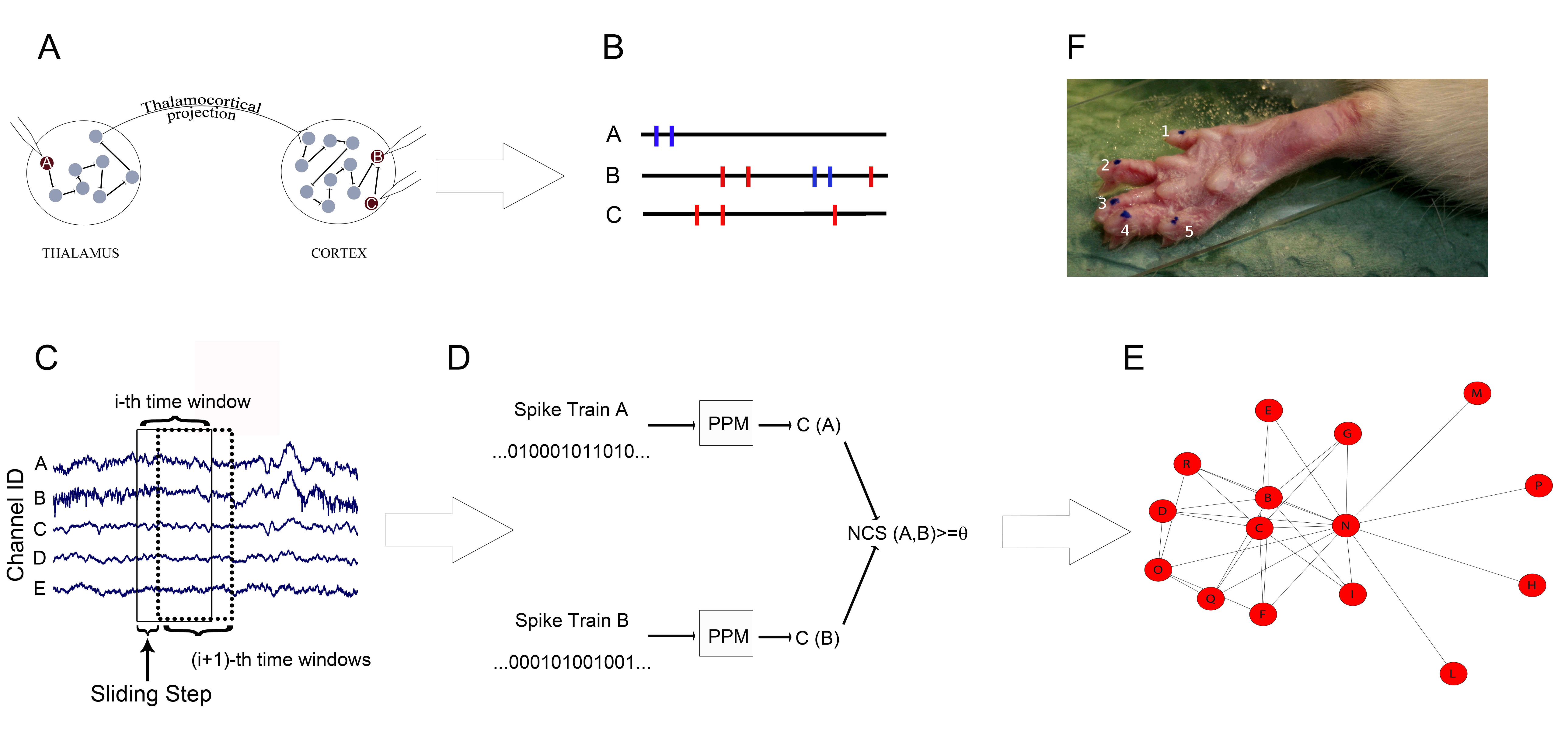}
\caption{
{\bf The proposed framework for the estimation of neuronal functional connectivity.}  (A) A recording session from thalamic and cortical regions. Arrows indicate the effective influence among neurons. The electrode tips record the neurons in dark red. (B) The firing patterns of the cortical neuron B produce common firing patterns both with neurons A and C but with different time delays. In particular, $C \rightarrow B$ (red spikes) can be easily inferred by correlation analysis instead of $A \rightarrow B$ (blue spikes) hardly detectable. (C) Recorded signals are processed in overlapping windows lasting hundreds of milliseconds. (D) Spike trains are modeled by VMMs and compressed by LCAs. The functional connectivity strength between the spike trains A and B is estimated by the length of the compressed spike trains (C(A), C(B)) used by the $NCS$ function. Whether $NCS(A,B)$ is greater than a fixed threshold then we can conclude that $A \rightarrow B$. (E) An example of functional graph extracted by recordings 
in the i$^{th}$ time window. (F) Typical sites of the rat paw for the tactile stimulation.
}
\label{fig1}
\end{figure}

\begin{figure}[!h]
\includegraphics[width=\textwidth]{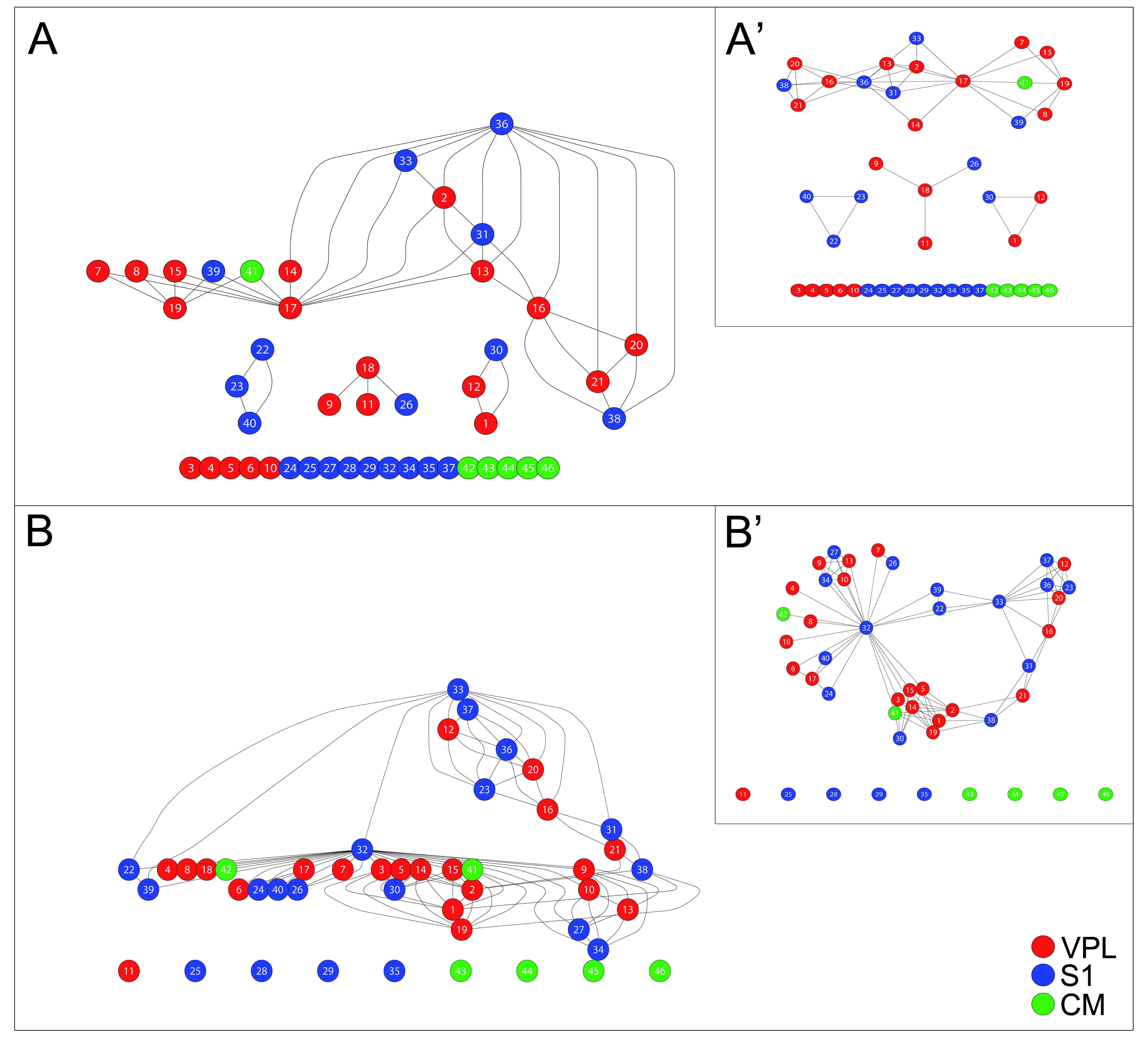}
\caption{
{\bf Example of stimulus evoked redistribution of spiking functional connections.} Functional weights are redistributed from the pre- (A) to post-stimulus (B) configurations. Red, blue and green nodes indicate neurons respectively from VPL, S1 and CM. Some neurons, not functionally connected in the pre-stimulus graph, are involved in the stimulus processing (3-6,10,24,28,32,34,37,42). Conversely, some neurons, previously employed, are excluded in the functional graph (11). Furthermore, many neurons change their functional roles. For instance, cortical neuron 32 is a hub node (node with high degree) with many functional connections with VPL and CM thalamic neurons in (B) while is not involved in (A). Again, cortical neuron 23 constitutes a small clique with cortical neurons 22 and 40 in (A) while in (B) becomes a small hub node with diverse thalamic and cortical neurons. 
}
\label{fig2}
\end{figure}

\begin{figure}[!h]
\includegraphics[scale=0.7]{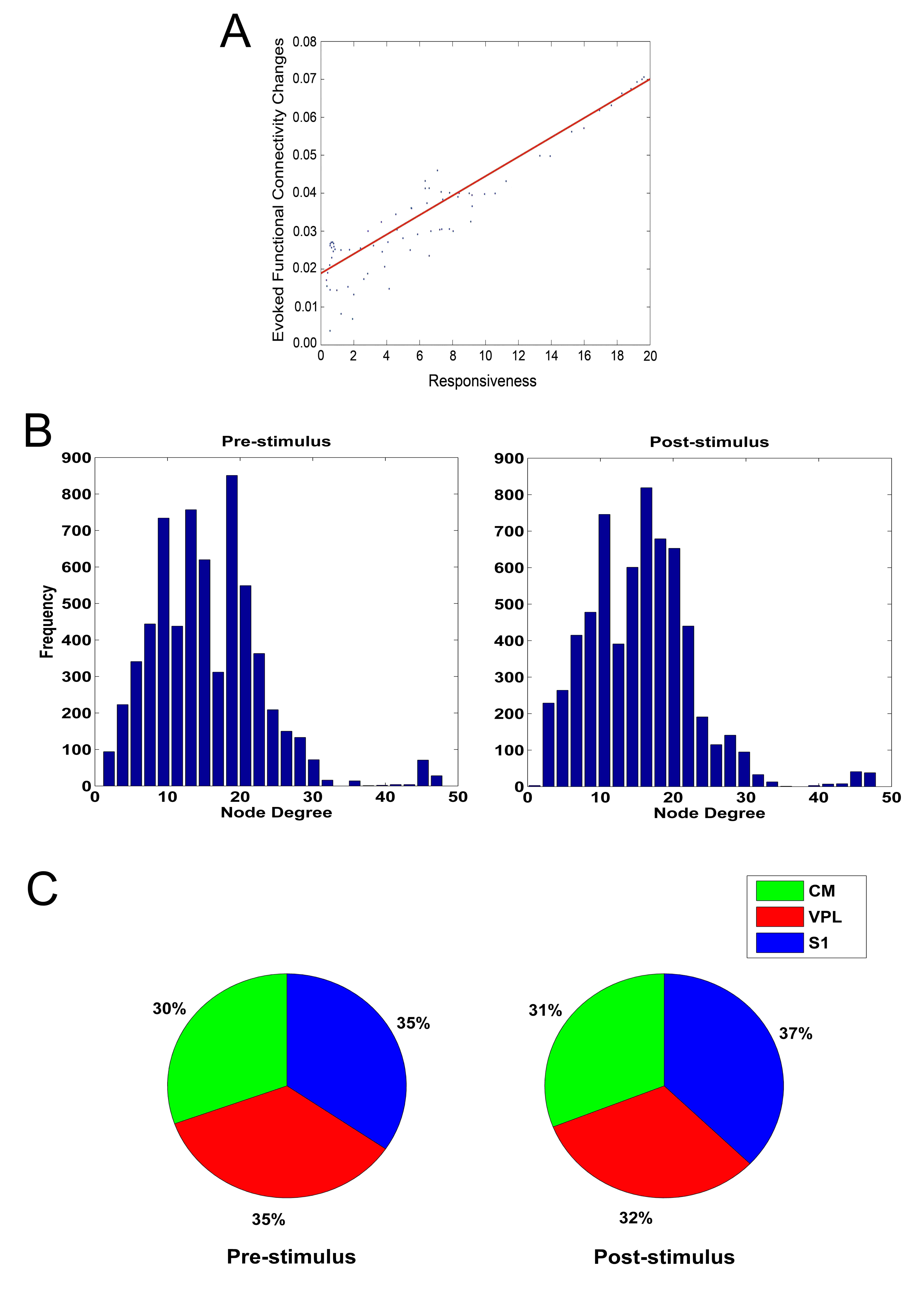}
\caption{
{\bf Salient facts of functional graphs in spiking activity.} (A) Correlation between the standard deviation of difference matrices versus the spike responsiveness in each stimulus session. Correlations were computed by least-square regression (red lines, R=0.637). (B) Average node degree distributions of functional graphs (spikes) extracted by pre- and post-stimulus conditions. (C) Average betweenness centrality balance over the three recorded regions (VPL, S1, CM) in both conditions.
}
\label{fig3}
\end{figure}

\begin{figure}[!h]
\includegraphics[width=\textwidth]{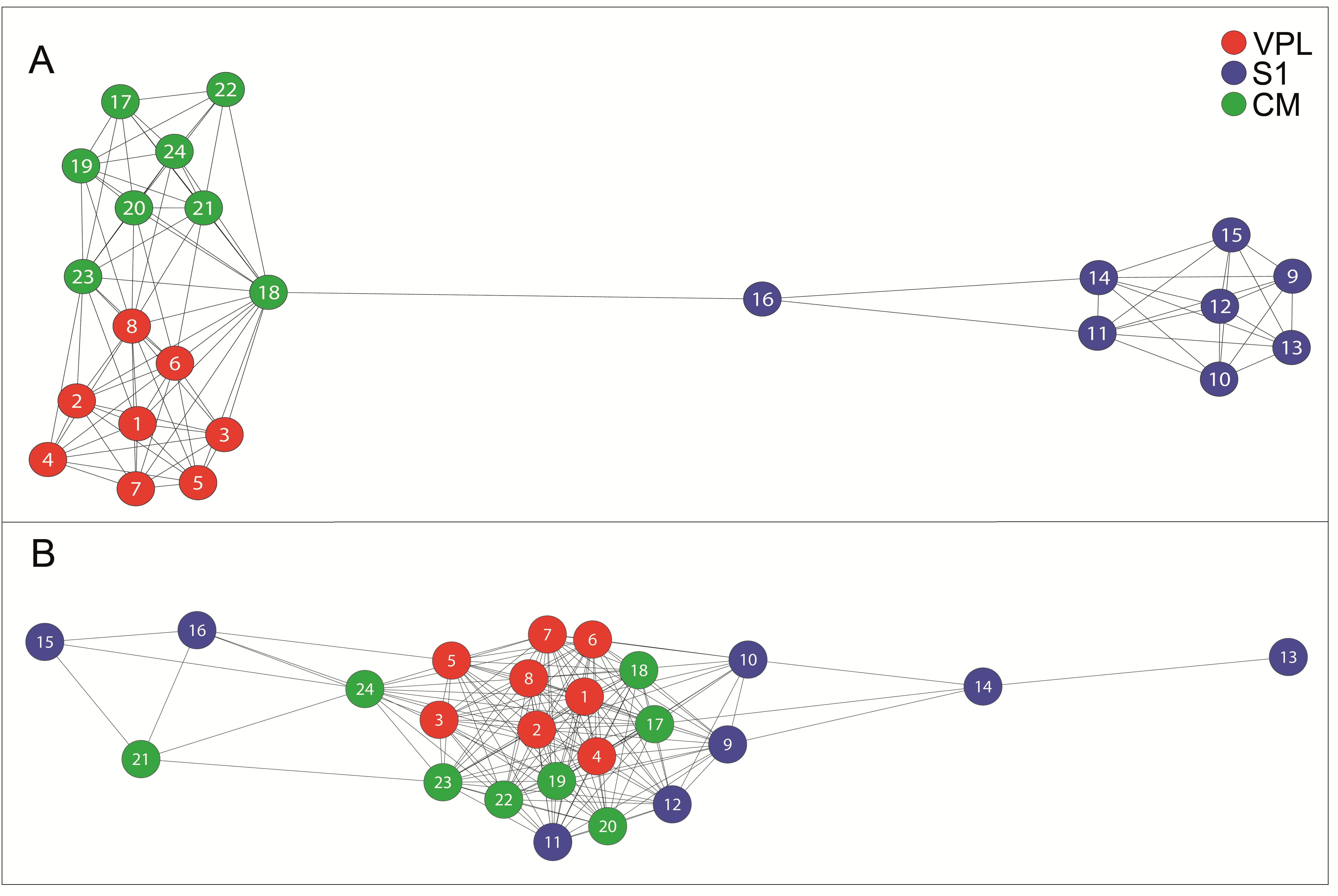}
\caption{
{\bf Example of LFP phases couplings. Functional connections are disposed on the LFP recording sites.} Green nodes represent CM channels, red nodes represent VPL channels and blue nodes represent S1 channels. (A) Before a tactile stimulation, LFPs are tightly coupled among the LFP channel of the same brain areas. (B) After an effective tactile stimulation, LFPs broke their inter-site synchronies and established cross-site phase couplings.
}
\label{fig4}
\end{figure}

\begin{figure}[!h]
\includegraphics[scale=0.7]{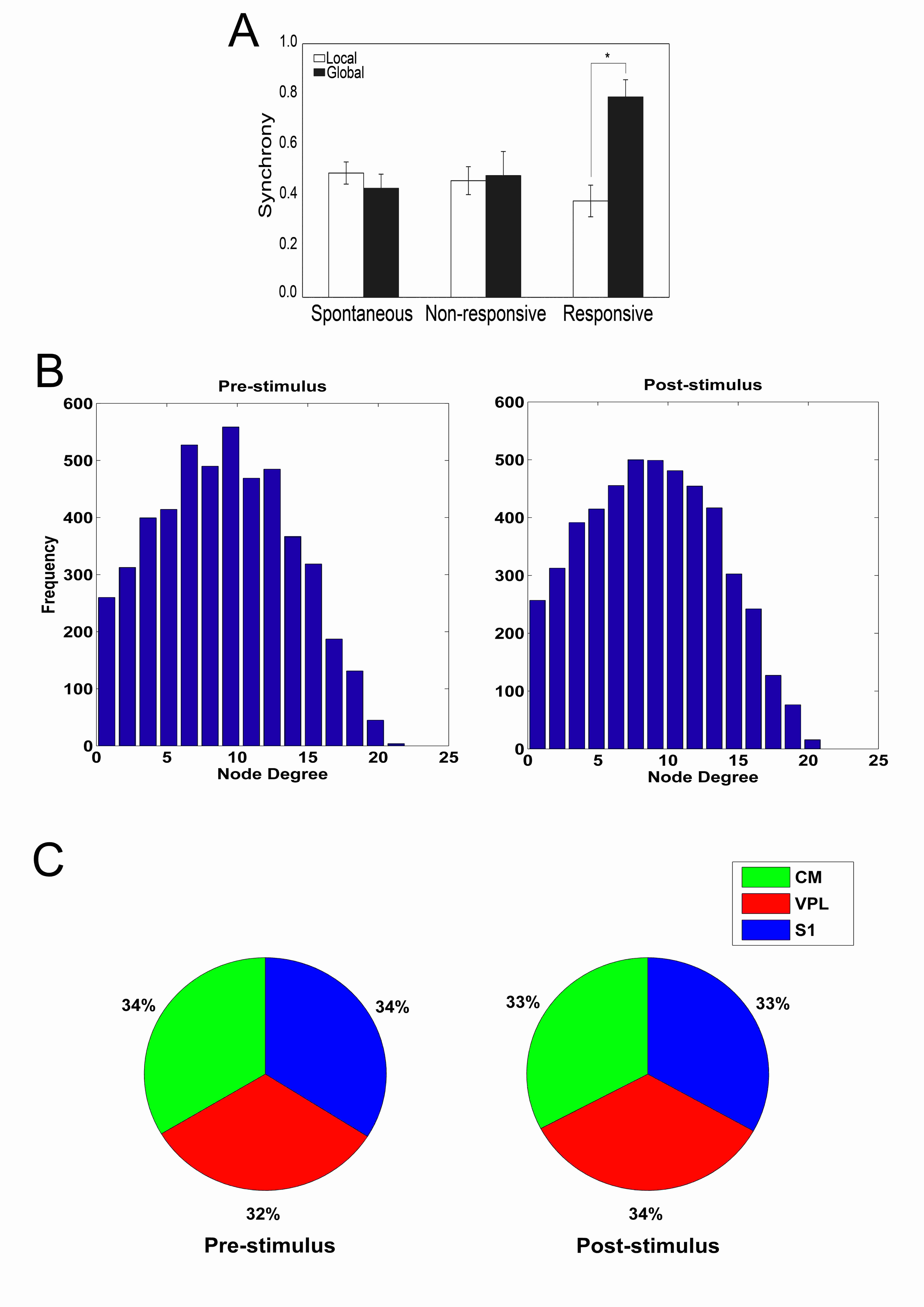}
\caption{
{\bf Salient facts of functional graphs in LFP.} (A) Synchrony of LFP phases in each recording site during spontaneous, spike responsive and spike non-responsive configurations. Responsive stimuli increase the global and decrease the local synchronies. (B) Average node degree distributions of functional graphs (LFPs) extracted by pre- and post-stimulus conditions . (C) Average betweenness centrality balance over the three recorded regions (VPL, S1, CM) in both conditions.
}
\label{fig5}
\end{figure}

\begin{figure}[!h]
\includegraphics[width=\textwidth]{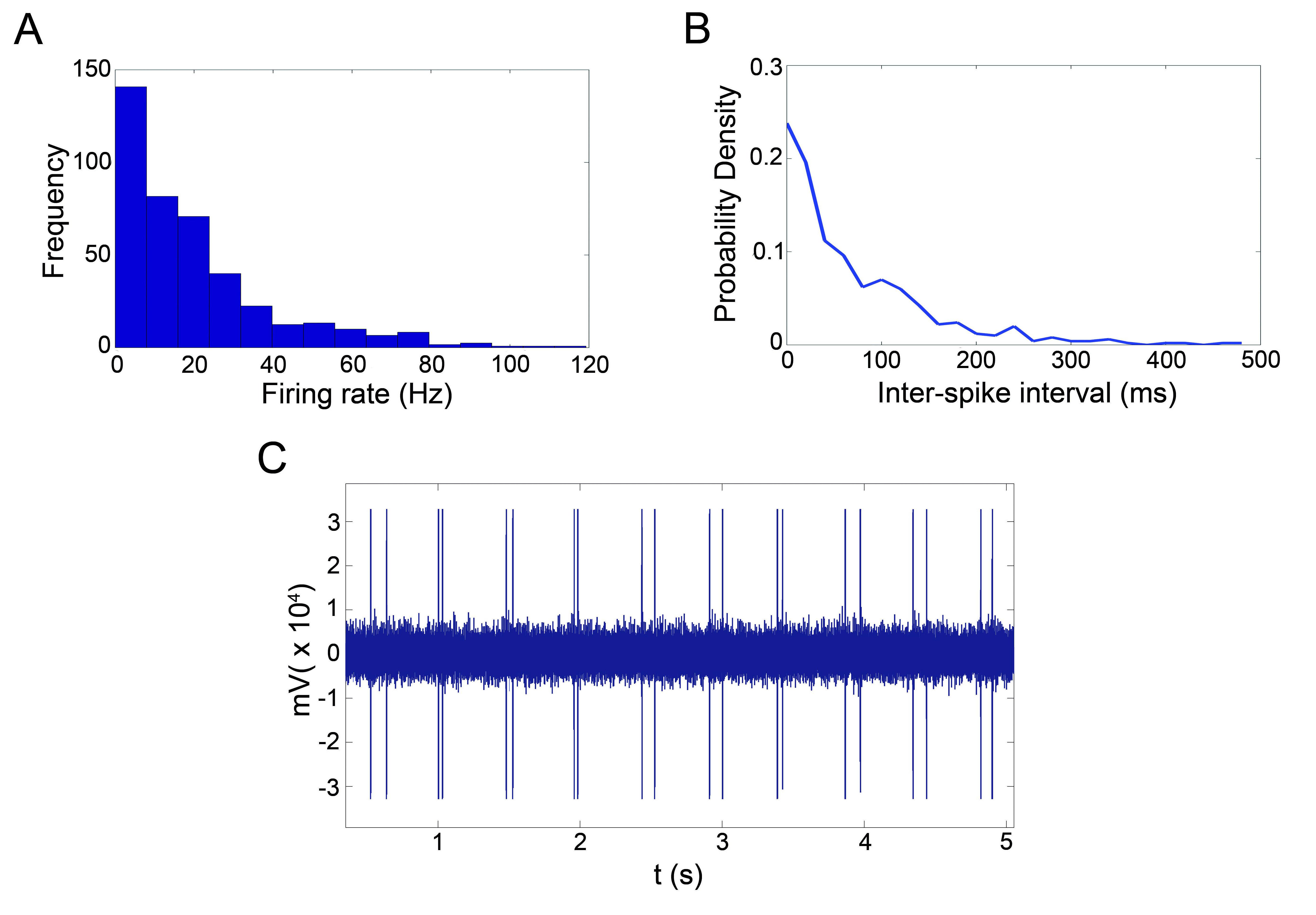}
\caption{
{\bf Basilar neurophysiological data.}  (A) Distributions of firing rate and (B) of Inter-spike Interval for the representative neurons. (C) Pattern of tactile stimulations by Arduino microcontroller.
}
\label{fig6}
\end{figure}

\begin{figure}[!h]
\includegraphics[width=\textwidth]{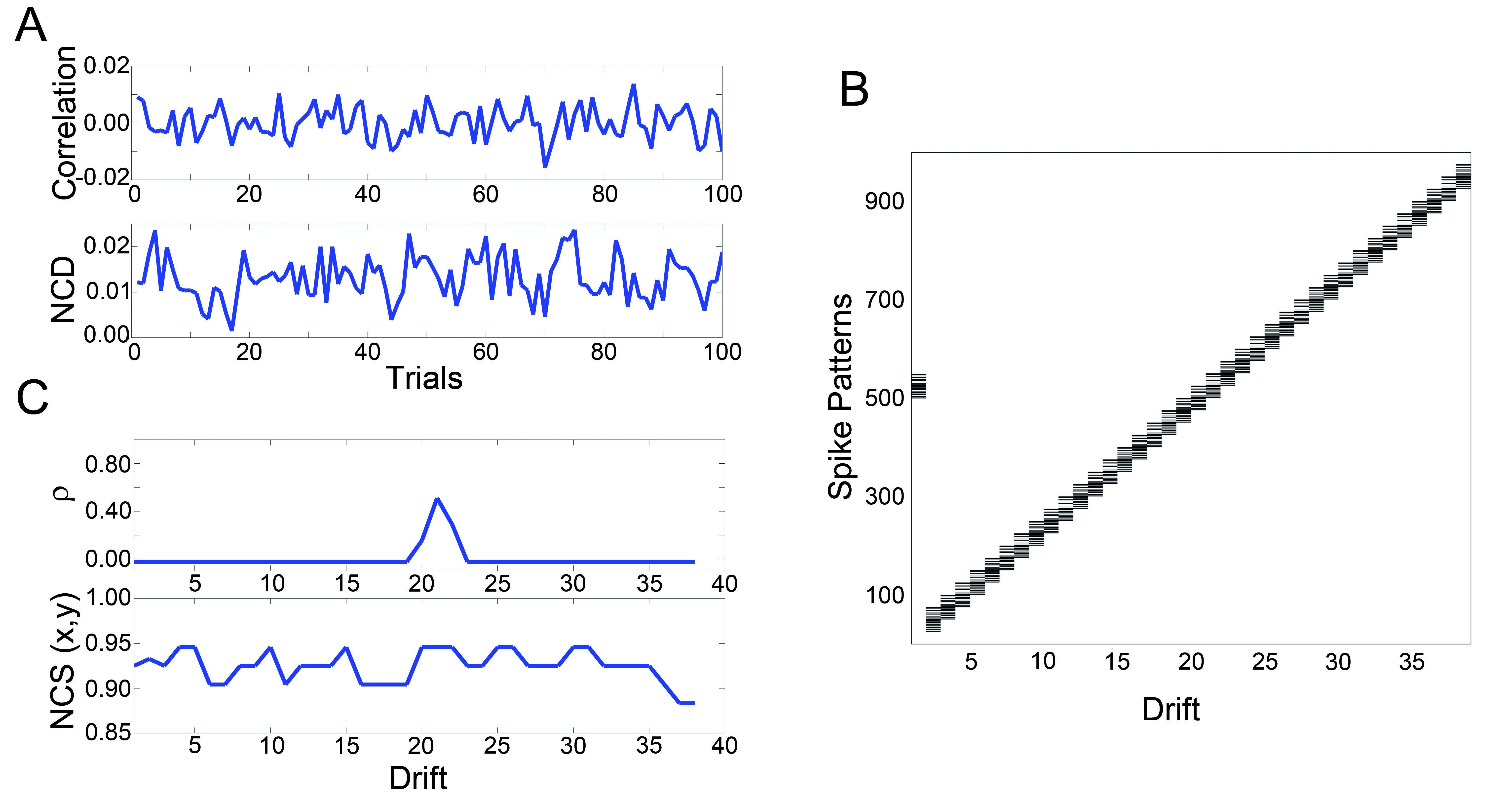}
\caption{
{\bf Efficacy of Normalized Compression Similarity (NCS) to detect long-range spike interactions.} (A) NCS and Pearson coefficient exerted on 100 couples of independent uniformly distributed binary sequences (1000 bits). Both functions do not show bias. (B) To test the capacity of NCS to detect significant interactions within time windows of 1000 ms, we fixed in the center of the first time-window a binary random pattern (50 ms long). The same pattern was replicated in a sequence of 36 time-windows drifting it from the initial to the window end. (C) NCS and Pearson coefficient were evaluated on these sequences: NCS is able to detect the interaction along the entire drifting process while Pearson coefficient is able to return significance only when the reference pattern is almost aligned in both sequences (drifts number 20-23).
}
\label{fig7}
\end{figure}

Interactions between neurons can generate very complex, time-delayed and asymmetric patterns. This represents a potential problem in experimental configurations where the communications between distant neurons are taken into consideration. Standard techniques like correlation analysis are, in many cases, unable to detect such events. Indeed dependencies between neurons from thalamus and cortex can last tens of milliseconds \cite{panzeri,aguilar,latham,storchi}. In a toy example (shown in Figure \ref{fig1}A-B) it is assumed the existence of the interaction between neurons A and B (A $\rightarrow$ B). The neuron A is recorded and its activation triggers the long chain that, from site A, produces firing activity to neuron B in another brain region. The direct path between neurons C and B allows, instead, for synchronous spike patterns well detectable by correlation analysis (Fig. \ref{fig1}B).\\
In general, in simultaneous recordings from separated brain sites, it is unlikely to find couples of physically wired neurons although axonal projections connect the sites. It's definitely more probable that spiking activity relations may reflect a complex anatomical substrate, where chains of activations exist between them provoking the observed coherent activity. Such a problem can be solved by mathematical tools able to model arbitrarily long temporal relationships.
In this work, we proposed a novel framework wherein spike trains with arbitrarily long temporal dependencies are modeled by Markov stochastic models. Typically, in regular Markov models, each state depends only on the previous state while higher-order Markov models suffer from high state-space computational complexity \cite{zaki}. Here, we used the Variable-order Markov Models (VMMs) \cite{buhlmann,begleiter} because they are able to overcome these limitations. 
Lossless compression algorithms (LCAs) represent one of the most efficient techniques to estimate VMMs \cite{begleiter}. Within the set of LCAs we chose the Prediction by Partial Matching (PPM) algorithm \cite{cleary,teahan} which is considered the best match between prediction accuracy and speed \cite{begleiter}. The last step consists to build a similarity function for this kind of spike train stochastic models.\\
In the last 15 years, Vitanyi and colleagues have developed a function, the Normalized Compression Distance ($NCD$) \cite{li,bennett,chen,cilibrasi}, which estimates the distance between symbolic sequences. This function performs the estimation directly by the sizes of the compressed sequences. In fact, can be proved that the better is the VMM estimation the shorter is the compressed sequence size. In this work, we redefined the $NCD$ function in order to reverse its assigned values pointing to similarity instead of distance. We called such function the Normalized Compression Similarity ($NCS$). Formally, given that $x$ and $y$ are two neural sequences (e.g. spike trains), the $NCS$ is defined as follows:
\begin{equation}
NCS(x,y) = 1 - NCD(x,y) = \frac{C(x\cdot y) - \min\{C(x),C(y)\}}{\max\{C(x),C(y)\}}
\end{equation}
where the $C$ function represents the compressed sequence length and $\cdot$ is the sequence concatenation operator (e.g. $0101\cdot101 = 0101101$). If $NCS(x,y)$ is close to $1$, the sequences $x$ and $y$ are considered similar. If close to $0$, the sequences are strongly dissimilar.\\ 
We therefore evaluated the $NCS$ function on time windows ($50$-$1000$ ms) of the recorded (binary) spiking activity (Fig. \ref{fig1}C-D) assuming that relative high values of similarity corresponded to actual functional connections (Fig. \ref{fig1}E). To note that significant $NCS$ values do not imply significant correlations but the opposite is true. Although the $NCS$ is asymmetric function, it is not relevant for the causal interaction analysis and consequently for effective connectivity.
Because $NCS$ has never been used as method to estimate the functional connectivity in neurophysiological data (although many LCAs has been used as information measure of synaptic efficacy \cite{london} and spike train similarities \cite{szczepanski,zippo,amigo}), our aim was to prove that $NCS$ effectively captures similar asynchronous spike patterns in comparison to correlation coefficient. Furthermore, we demanded that $NCS$ did not bias its estimations with respect to independent spike trains generated by simulations.\\
To address these requirements we performed two experiments: i) we evaluated the $NCS$ and the Pearson correlation coefficient over a set of independent binary spike trains and ii) we evaluated the $NCS$ and the correlation coefficient with couples of binary spike trains ($1000$ ms long) containing an identical, short and random spike pattern ($50$ ms) that is fixed and centered in the first train and drifts (from left to right) in the second trains (see Fig. \ref{fig7}B, overall $37$ drifts). The latter procedure creates synthetic spike trains wherein the common pattern  is increasingly distant.\\
In the first experiment we found that the NCS method did not show any significance for independent spike trains as well as for the Pearson coefficient ($\mu = -1.6\cdot 10^{-5}$, $\sigma = 0.005$ for Pearson coefficient and $\mu = 0.013$, $\sigma = 0.004$ for $NCS$, Fig. \ref{fig7}A). This guarantees that the $NCS$ function is unbiased to independent distributed binary spike trains. In the second experiment, the $NCS$ proved its efficacy to detect long-range interactions where Pearson coefficient fails. In fact, during the whole shifting epochs, significance held while Pearson coefficient showed significant only when the two spike patterns were very spatially close (Fig. \ref{fig7}C, drift number 18-23).

\subsection*{Functional connectivity by LFP phase synchrony}
LFPs are low frequency signals reflecting a wide range of synaptic events. In this work we investigated the synchrony of LFP phases originated in different recording sites during spontaneous and tactile evoked activities. We measured phase synchronies between two recorded LFP sequences ($x$ and $y$) by the following function
\begin{equation}
\gamma(x,y) = \left|\left< e^{i (arg(H(x))-arg(H(y)))} \right>\right|
\end{equation}
where $e$ is Napier's constant, $H$ is Hilbert Transform, $arg$ is the argument function and $i$ is the imaginary unit. The Hilbert transform and the argument were computed with, respectively, the \texttt{hilbert} and the \texttt{angle} Matlab functions \cite{pereda,montemurro}. When $\gamma(x,y)$ is equal to $1$ ($0$), then $x$ and $y$ are perfectly synchronous (asynchronous).

%\subsection*{Spike responsiveness}
%We related different aspects emerging from the spiking activities in response to tactile stimuli. In particular, since these activities showed deep differences among the selected paw sites in the different experiments, we decided to compare them with their specific functional connectivity data, separating spike and LFP responsiveness. Peri-Stimulus Time Histograms (PSTHs), with 5 ms binning, showed spike responsiveness in 50 ms time windows before and after tactile stimuli. Spike responsiveness was obtained subtracting the PSTH of spiking activity before the stimulus from the PSTH after the stimulus.

\subsection*{Complex Brain Network}
By using the NCS and $\gamma$ functions, we estimated the functional connectivity of the recorded neuronal networks. We first split each recorded sequence into equal-length time windows (Fig. \ref{fig1}) and then we computed the adjacency matrix for all neurons or electrodes. The resulting matrices exhibited values in the unitary interval. We repeated the analyses with different window sizes from $50$ ms to $1$ s (fixing the maximum dependency order to half of the window size). The functional connectivity extracted from extracellular recordings can be represented by graphs.\\ 
A variable threshold (typically equal to a higher percentile of the weight distribution and vary between $0.2$ and $0.8$) selected the strongest connections, thus allowing for the construction of the functional connectivity graphs. The choice of a threshold is related to the window length used to estimate the connectivity strength. Larger windows produce less connections and to keep the network sparsity quite constant, smaller thresholds must be chosen.\\
For the analysis of these graphs, we introduced a set of common statistics from the Complex Network Theory able to detect possible matches between the extracted graphs and prominent topologies like small-world networks. A small-world network is generally obtained by evolving a basic ring lattice graph, where each node is connected to their $K$ neighbors. The chosen neighborhood involves typically much less nodes than the total node number $N$ ($N\gg K\gg \ln(N)\gg1$). The graph evolution is achieved by randomly adding and removing edges from the starting graph \cite{watts,barabasi}. The resulting graph has many, typically small, quasi-complete subgraphs (cliques) where each node is connected to every other node in the clique. Furthermore, small-world networks exhibit short average distances between nodes.\\
From a functional perspective, small-world networks can express two important information processing features: information integration and segregation \cite{tononi,bullmore}. Functional segregation recruits specialized processing within densely interconnected nodes (cliques). Functional integration combines information processed in distributed nodes or cliques. These network properties can be measured by two statistics: the clustering coefficient ($C$) and the characteristic path length ($L$) \cite{watts,rubinov}. The former measures how close the neighbors of a node are to being a clique. The latter estimates the average shortest path length in the graph, i.e. how much the nodes are accessible. Both measures, implemented in a Matlab toolbox \cite{rubinov}, were used for our network analyses (\texttt{clustering\_coef\_bu.m}, \texttt{charpath.m}).\\
In complex network theory, several graph measures take specific meaning only if they are compared to the same graphs subject to randomization or latticization (often called null networks) \cite{rubinov}. Both procedures guarantee that the node degree distributions of the original graphs were preserved. We computed, by using the Matlab function \texttt{randmio\_und.m}, the randomized version of our graphs and we estimated $C^r$ and $L^r$ ($C^l$ by latticization, \texttt{latmio\_und.m}). These null network values are required to verify the small-world nature of the graphs. In fact, classical and novel measures of small-worldness such as $S=\frac{C/C^r}{L/L^r}$ \cite{humphries2008} and $\omega=\frac{L^r}{L} - \frac{C}{C^l}$ \cite{telesford} state that, respectively, if $S > 1$ or $\omega$ is close to $0$, the graph can be considered a small-world network.
The functional graphs obtained by our analysis were further characterized to study the information flow. For this aim, we computed a measure of centrality (betweenness) for graph nodes \cite{kintali}, an estimate of the number of shortest paths from all vertices to all others that pass through that node. Because it can be interpreted as a measure of the load of a node within the network \cite{rubinov}, the distribution of node centrality highlights how the information flow is balanced within graphs.\\
For the same purpose we further studied the community structure of our graphs. Communities emerge from graphs by applying a clustering algorithm to nodes \cite{newman}. In this work, communities represent the aggregated functional units under investigation and by counting the number of them we can understand how node graphs are aggregated in each experimental condition. Ultimately, we analyzed networks that evolved in time dropping and recruiting nodes and connections and networks from different experimental conditions. Such a methodology requires the discussion of potential issues \cite{maslov,zalesky}.\\ 
First, unconnected nodes are rarely but can occur after adjacency matrices were binarized. For this reason, we removed graphs in which less than the 99\% of nodes are connected. Second, network statistics were applied on network with different sizes (for spiking activity) because recording sessions return a variable number of active neurons. However, by analyzing the observed variance of network size we concluded that $C$ and $L$ cannot be significantly affected by our network size changes. Significant changes appear for synthetic networks that increased their size by orders of magnitude. However, we discarded graphs that are outliers (beyond 5$^{th}$ and 95$^{th}$ percentile) of the node, edge and density number distributions in order to obtain a better homogeneity.          

\subsection*{Visualization of graphs}
Graphs are visualized by using the function \texttt{LayeredGraphPlot} of Mathematica (Wolfram Research, Inc., Mathematica, Version 8.0, Champaign, IL). This function implements the Sugiyama algorithm \cite{sugiyama}. A Layer layout helps to understand the information flow in the network and the specific roles performed by each node.
Graphs in the insets are visualized by using the function \texttt{GraphPlot} of Mathematica that implements the spring-electrical algorithm \cite{fruchterman} that is typically used to draw small-world networks.

\subsection*{Computational Model}
\begin{figure}[!h]
\includegraphics[scale=0.75]{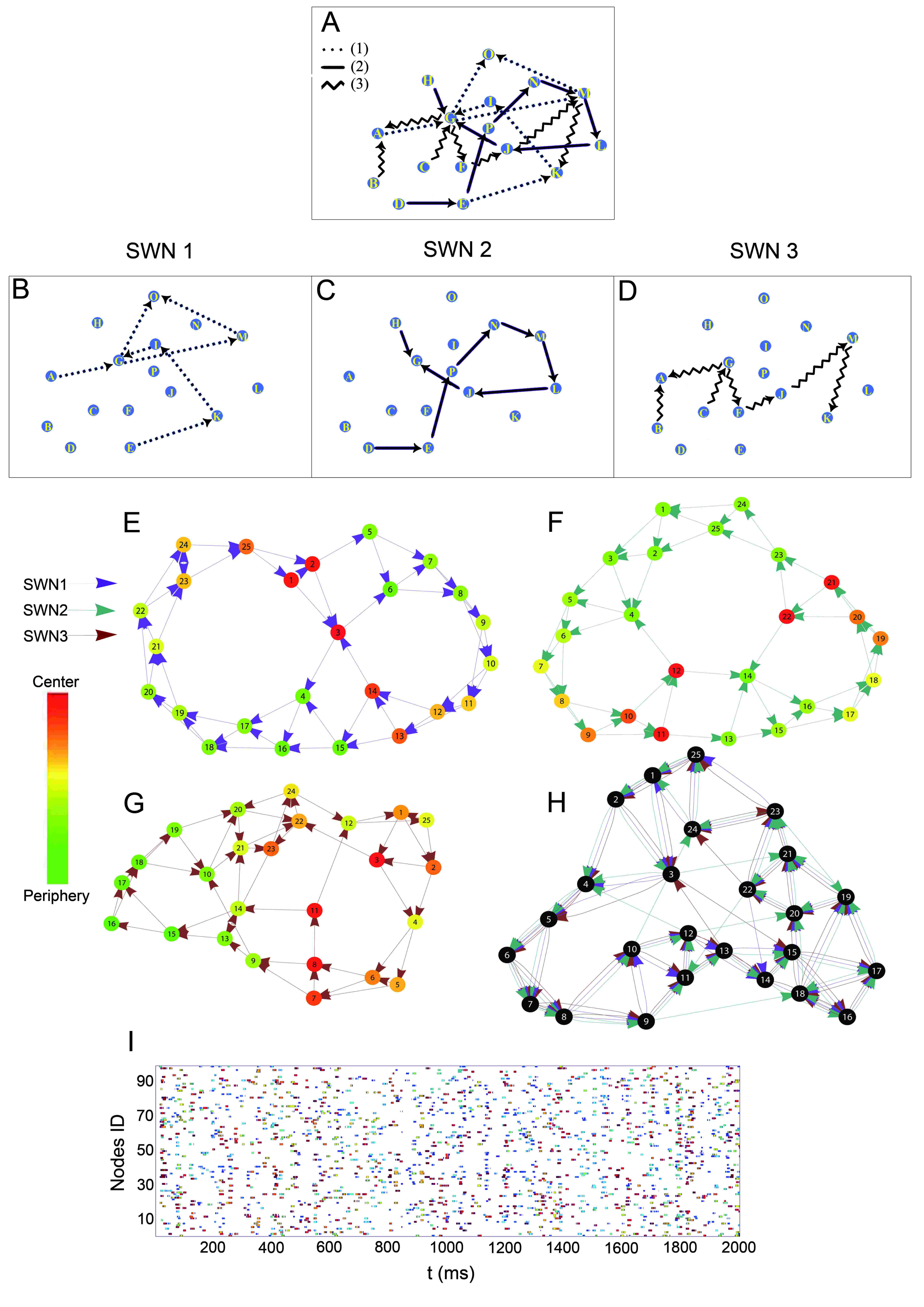}
\caption{
{\bf Fundamental assumptions of the computational model.} A toy example (16 nodes) showing three small-world networks (B-D) embedded into one network (A). The firing criterion is illustrated in graphs (E-H): each small-world network gives a specific (betweenness) centrality in its nodes. Nodes with low values of centrality (from periphery) spike first, nodes from center, later. (I) A sample of synthetic spiking activity from 100 nodes. Colors are specific for each small-world network.
}
\label{fig8}
\end{figure}

So far, Hebb's idea of {\it cell assemblies} represents a fundamental theory supporting important physiological events. Guided by results from recordings, we hypothesized that a cell assembly can be functionally organized as a small-world network and we investigated such hypothesis by computer simulations. As originally thought, a cell assembly can be constituted by groups of neurons (even anatomically dispersed) and the membership to each assembly can be dynamical \cite{hebb}.\\
Specifically, we brought back such considerations in a computational model to investigate two hypotheses: (i) Can a network model based on previous assumptions can be consistent with the observed experimental facts? (ii) assuming the previous as true, how many assemblies (small-world networks) can exist over a set of neurons keeping consistency with observations? \\
Typical simulation frameworks require a choice of neuron models (Integrate and Fire, Izhikevich, Hodgkin and Huxley, etc.) and of a defined network layout (nodes and connections). These choices can be very crucial and become even more important if the aim of the study is the functional organization of the units. For this reason, we proceeded following an unconventional approach assuming that cell assemblies are effectively functionally organized as small-world networks on large-scale networks and, sampling the activity of a random subset of nodes, we wondered if such activity can elicit small-world organization as well.\\ 
So we first assumed that a set of small-world networks exists over a set of available nodes. If we consider each small-world network as a cell assembly \cite{hebb}, the small-world networks of our model can share their nodes.\\
As a whole these facts define the structural property of the model and can be summarized as follow:
\begin{enumerate}
\item Each neuron is represented by a node.
\item Functional connections between neurons are represented by edges.
\item Neurons are functionally organized as small-world networks.
\item Many small-world networks are embedded within the simulated network thus each node may belong to more than one small-world topologies (Fig. \ref{fig8}A-D). 
\item According to the typical amount of neocortical neurons in a microcolumn \cite{mountcastle}, the sizes of small-world networks are randomly sampled by a uniform distribution $\sim U(50, N/100)$ where $N$ is the number of nodes.
\end{enumerate}
From a formal point of view, the network structure can be interpreted by a Multigraph $G=(V,E)$, the set $V$ represents the nodes and the multiset $E$ represents the unordered list of edges inherited by each small-world network.\\ 
To establish the node functioning we chose a primary criterion claiming that brain processing takes place by functional segregation and integration \cite{tononi}. This concept, supported by many experimental evidences, explains how segregated specific brain areas work together to produce globally integrated behaviors, cognitions and percepts. We noted that small-world networks and integration-segregation criterion are closely related. Computations performed in (specialized) peripheral nodes, can be subsequently integrated in central nodes by virtue of the small average shortest path length of small-world networks. Indeed, it can be easily shown, by using simple computer simulations, that nodes in small-world networks had a wide spectrum of centrality values. This had a heavy-tailed distribution in contrast to random networks distributed as a Gaussian-like distribution, i.e. every node had almost the same centrality.\\
In order to implements these criteria in our model, nodes fires following a rank-order dictated by the centrality values estimated by using the centrality. The centrality of nodes can be estimated by several measures (e.g. betweenness, pagerank, node degree, etc.).\\
As a whole, these last facts can be summarized as follow:
\begin{enumerate}
\item Each small-world network represents the processing of a specific information (the working hypothesis).
\item At most two randomly chosen small-world networks are allowed to run independently in a time window. A single neuron can work either alternately or concurrently within different networks of pre and post-synaptic nodes. This assumption claims that neurons are not exclusive and that they can partake simultaneously to many (up to 2) tasks. 
\item Neurons express their spikes within 10 ms time windows. Neurons fire following a centrality criterion based on the node betweenness centrality. A neuron with high betweenness fires its spikes after a lower betweenness neuron. This dynamic picture respects the information integration-segregation paradigm (Fig. \ref{fig8}E-H) \cite{tononi}. The choice of the window length is arbitrarily and is proportional to the firing rate.
\item Simulation timesteps are set to 1 millisecond. Spike propagation times are uniformly distributed within the range $[1,3]$.
\end{enumerate}
The algorithm governing this network is as follows:

\begin{algorithm}[H]
 \KwIn{$n\_swn$, $n\_nodes$, timestep;}
 \KwOut{the binary $n\_nodes$-by-$timesteps$ matrix $trains$}
 $p \leftarrow 0.05$\;
 $k \leftarrow 5$\;
 \For{$i\leftarrow 1$ \KwTo $n\_swn$}{
  $n \leftarrow $ \texttt{subsample($n\_nodes$)}\;
  $G \leftarrow $ \texttt{random\_watts\_strogatz($n,k,p$)}\;
  $wins \leftarrow $ \texttt{computeWindows($G$)}\;
  \ForEach{time windows in $wins$}{
    \If{$0.01 < $ \texttt{rand()}}{
	$centrality \leftarrow $ \texttt{BetweennessCentrality($G$)}\;
	\ForEach{node $A$ in $centrality$}{
	  $trains[A,wins + centrality(A)] \leftarrow 1$\;
	    \ForEach{output node $B$ of $A$}{
	      $trains[A,wins + centrality(A) + centrality(B)] \leftarrow 1$\;
	    }
	}
      }
    }
 }
  
\end{algorithm}
\noindent where the function \texttt{subsample()} selects a subset of nodes out of the set of available ones. The function \texttt{random\_watts\_strogatz()} returns a random graphs built following the Watts-Strogatz algorithm, the function \texttt{computeWindows()} computes the length of the execution windows for the current graph and returns the list of such windows. The function \texttt{rand()} returns a uniformly generated random number between $0$ and $1$. At last, the function \texttt{BetweennessCentrality()} computes and returns the centrality of each node.\\
The model has been developed using the Python environment \cite{lutz}. For the generation of small-word networks we used the networkx package (available at http://networkx.lanl.gov/). Fig. \ref{fig8}I shows a raster plot obtained by sampling the simulated activity. The darkest blue represents the no-spike event and the other spike colors are associated with the specific small-word topologies that generated them. 
The algorithm describes a core loop where each small-world network is first randomly created by the library routine \texttt{watts\_strogatz\_graph} (probability of rewiring equal to $0.05$). Small-world networks take only a fraction of the total node number and several small-world networks can share a subset of their nodes. In a second stage, the generated small-world network expresses its spiking activity following the betweenness centrality as in point 2 of the dynamical assumption. Low level uniformly distributed noise is added to the spike propagation time. The spikes of the current network occur randomly in equal size time windows (10 timesteps). The spike activity is saved and the loop restarts (source codes are available at http://code.google.com/p/swn-neuronal-networks/).

% Do NOT remove this, even if you are not including acknowledgments
\section*{Acknowledgments}
We wish to thank M.E. Giardini, L. Bonini and C. Toschi for helpful suggestions. Finally, we thank Katherine Davis for the revision of the english. 

%\section*{References}
% The bibtex filename

\bibliography{template}

\section*{Figure Legends}

\pagebreak

\section*{Tables}

\begin{table}[!h]
\centering
\begin{scriptsize}
\begin{tabular}{|c|c|c|c|c|c|c|c|c|c|c|}
\hline
%{\bf \#nodes} & {\bf \#edges} & {\bf win (ms)} & $\theta$ & $C$ & $C^{l}$ & $C^r$ & $L$ & $L^r$ & $S = \frac{C/C^r}{L/L^r}$ & $\omega = \frac{L^r}{L} - \frac{C}{C^{l}}$ \\ \hline
nodes & edges & win & $\theta$ & $C$ & $C^{l}$ & $C^r$ & $L$ & $L^r$ & $S$ & $\omega$ \\ \hline
$57 \pm 14$ & $1117 \pm 385$ & 50 & 0.5 & $0.91 \pm 0.08$ & $0.92 \pm 0.08$ & $0.59 \pm 0.08$ & $1.48 \pm 0.11$ & $1.48 \pm 0.11$ & $1.51 \pm 0.00$ & $0.01 \pm 0.00$ \\ \hline
$57 \pm 14$ & $947  \pm 317$ & 50 & 0.6 & $0.89 \pm 0.09$ & $0.95 \pm 0.08$ & $0.61 \pm 0.08$ & $1.54 \pm 0.10$ & $1.55 \pm 0.09$ & $1.45 \pm 0.01$ & $0.08 \pm 0.01$ \\ \hline
$57 \pm 14$ & $1516 \pm 216$ & 250 & 0.3 & $0.92 \pm 0.02$ & $0.92 \pm 0.01$ & $0.59 \pm 0.02$ & $1.36 \pm 0.14$ & $1.31 \pm 0.06$ & $1.50 \pm 0.00$ & $-0.03 \pm 0.01$\\ \hline
$57 \pm 14$ & $1311 \pm 246$ & 250 & 0.4 & $0.90 \pm 0.04$ & $0.96 \pm 0.03$ & $0.61 \pm 0.03$ & $1.36 \pm 0.12$ & $1.14 \pm 0.07$ & $1.24 \pm 0.05$ & $-0.09 \pm 0.01$ \\ \hline
$57 \pm 14$ & $1294 \pm 228$ & 500 & 0.3 & $0.95 \pm 0.05$ & $0.98 \pm 0.04$ & $0.54 \pm 0.05$ & $1.51 \pm 0.38$ & $1.47 \pm 0.08$ & $1.70 \pm 0.04$ & $0.00 \pm 0.01$ \\ \hline
$57 \pm 14$ & $1176 \pm 236$ & 500 & 0.4 & $0.94 \pm 0.04$ & $0.97 \pm 0.04$ & $0.57 \pm 0.06$ & $1.56 \pm 0.35$ & $1.49 \pm 0.08$ & $1.55 \pm 0.02$ & $-0.01 \pm 0.02$ \\ \hline
$57 \pm 14$ & $1198 \pm 170$ & 1000 & 0.2 & $0.91 \pm 0.01$ & $0.86 \pm 0.03$ & $0.53 \pm 0.06$ & $1.31 \pm 0.06$ & $1.27 \pm 0.05$ & $1.68 \pm 0.00$ & $-0.08 \pm 0.00$ \\ \hline
$57 \pm 14$ & $876 \pm 164$ & 1000 & 0.3 & $0.89 \pm 0.05$ & $1.00 \pm 0.04$ & $0.50 \pm 0.09$ & $1.32 \pm 0.08$ & $1.34 \pm 0.06$ & $1.81 \pm 0.06$ & $0.11 \pm 0.00$\\ \hline
\end{tabular}
\end{scriptsize}
\caption{Network statistics for spontaneous spiking activity.}\label{table1}
\end{table}

\begin{table}[!h]
\centering
\begin{scriptsize}
\begin{tabular}{|c|c|c|c|c|c|c|c|c|c|c|}
\hline
\multicolumn{11}{|c|}{ pre-stimulus }\\ \hline
%\#nodes & \#edges & win (ms) & $\theta$ & $C$ & $C^{l}$ & $C^r$ & $L$ & $L^r$ & $S = \frac{C/C^r}{L/L^r}$ & $\omega = \frac{L^r}{L} - \frac{C}{C^{l}}$ \\ \hline
nodes & edges & win & $\theta$ & $C$ & $C^{l}$ & $C^r$ & $L$ & $L^r$ & $S$ & $\omega$ \\ \hline
$56 \pm 17$ & $889 \pm 297$ & 100 & 0.6 & $0.86 \pm 0.08$ & $0.98 \pm 0.05$ & $0.62 \pm 0.04$ & $1.55 \pm 0.11$ & $1.59 \pm 0.09$ & $1.42 \pm 0.09$ & $0.01 \pm 0.01$ \\ \hline
$56 \pm 17$ & $1161  \pm 364$ & 100 & 0.5 & $0.89 \pm 0.06$ & $0.93 \pm 0.04$ & $0.60 \pm 0.04$ & $1.47 \pm 0.12$ & $1.48 \pm 0.11$ & $1.49 \pm 0.03$ & $0.04 \pm 0.00$ \\ \hline
$56 \pm 17$ & $1356 \pm 439$ & 100 & 0.4 & $0.90 \pm 0.04$ & $0.91 \pm 0.04$ & $0.59 \pm 0.4$ & $1.40 \pm 0.13$ & $1.40 \pm 0.13$ & $1.52 \pm 0.01$ & $0.00 \pm 0.00$\\ \hline
$56 \pm 17$ & $1409 \pm 476$ & 100 & 0.3 & $0.91 \pm 0.04$ & $0.91 \pm 0.04$ & $0.59 \pm 0.04$ & $1.37 \pm 0.14$ & $1.37 \pm 0.14$ & $1.53 \pm 0.00$ & $0.00 \pm 0.00$ \\ \hline
\multicolumn{11}{|c|}{ post-stimulus }\\ \hline
$56 \pm 17$ & $897 \pm 302$ & 100 & 0.6 & $0.87 \pm 0.01$ & $0.97 \pm 0.08$ & $0.63 \pm 0.09$ & $1.55 \pm 0.11$ & $1.60 \pm 0.09$ & $1.41 \pm 0.01$ & $0.04 \pm 0.02$ \\ \hline
$56 \pm 17$ & $1159 \pm 391$ & 100 & 0.5 & $0.90 \pm 0.10$ & $0.93 \pm 0.09$ & $0.60 \pm 0.09$ & $1.47 \pm 0.12$ & $1.48 \pm 0.12$ & $1.49 \pm 0.00$ & $0.06 \pm 0.00$ \\ \hline
$56 \pm 17$ & $1371 \pm 480$ & 100 & 0.4 & $0.91 \pm 0.09$ & $0.91 \pm 0.09$ & $0.59 \pm 0.09$ & $1.39 \pm 0.14$ & $1.39 \pm 0.14$ & $1.53 \pm 0.00$ & $-0.00 \pm 0.00$ \\ \hline
$56 \pm 17$ & $1424 \pm 517$ & 100 & 0.3 & $0.91 \pm 0.09$ & $0.91 \pm 0.09$ & $0.59 \pm 0.09$ & $1.37 \pm 0.16$ & $1.37 \pm 0.16$ & $1.53 \pm 0.00$ & $0.00 \pm 0.00$\\ \hline
\end{tabular}
\end{scriptsize}
\caption{Network statistics for evoked spiking activity.}\label{table2}
\end{table}

\begin{table}[!h]
\centering
\begin{scriptsize}
\begin{tabular}{|c|c|c|c|c|c|c|c|c|c|c|}
\hline
%\#nodes & \#edges & win (ms) & $\theta$ & $C$ & $C^{l}$ & $C^r$ & $L$ & $L^r$ & $S = \frac{C/C^r}{L/L^r}$ & $\omega = \frac{L^r}{L} - \frac{C}{C^{l}}$ \\ \hline
nodes & edges & win & $\theta$ & $C$ & $C^{l}$ & $C^r$ & $L$ & $L^r$ & $S$ & $\omega$ \\ \hline
$24$ & $186 \pm 60$ & 50 & 0.5 & $0.87 \pm 0.05$ & $0.95 \pm 0.05$ & $0.48 \pm 0.12$ & $1.65 \pm 0.242$ & $1.53 \pm 0.14$ & $1.04 \pm 0.00$ & $0.00 \pm 0.01$ \\ \hline
$24$ & $116  \pm 46$ & 50 & 0.6 & $0.94 \pm 0.09$ & $0.85 \pm 0.12$ & $0.29 \pm 0.12$ & $2.06 \pm 0.41$ & $1.92 \pm 0.03$ & $1.25 \pm 0.04$ & $-0.12 \pm 0.05$ \\ \hline
$24$ & $159 \pm 37$ & 250 & 0.4 & $0.92 \pm 0.05$ & $0.93 \pm 0.04$ & $0.42 \pm 0.07$ & $1.83 \pm 0.25$ & $1.62 \pm 0.11$ & $1.17 \pm 0.10$ & $-0.09 \pm 0.01$\\ \hline
$24$ & $92 \pm 25$ & 250 & 0.5 & $0.92 \pm 0.09$ & $0.81 \pm 0.13$ & $0.23 \pm 0.08$ & $1.85 \pm 0.41$ & $2.13 \pm 0.29$ & $1.55 \pm 0.06$ & $0.02 \pm 0.03$ \\ \hline
$24$ & $198 \pm 32$ & 500 & 0.3 & $0.90 \pm 0.03$ & $0.95 \pm 0.03$ & $0.51 \pm 0.06$ & $1.62 \pm 0.01$ & $1.47 \pm 0.06$ & $1.068 \pm 0.00$ & $-0.04 \pm 0.01$ \\ \hline
$24$ & $151 \pm 25$ & 500 & 0.4 & $0.93 \pm 0.05$ & $0.94 \pm 0.04$ & $0.40 \pm 0.06$ & $1.84 \pm 0.23$ & $1.64 \pm 0.08$ & $1.12 \pm 0.02$ & $-0.10 \pm 0.02$ \\ \hline
$24$ & $113 \pm 12$ & 1000 & 0.2 & $0.92 \pm 0.07$ & $0.92 \pm 0.07$ & $0.31 \pm 0.05$ & $1.70 \pm 0.39$ & $1.87 \pm 0.07$ & $3.25 \pm 0.02$ & $0.09 \pm 0.07$ \\ \hline
$24$ & $87 \pm 11$ & 1000 & 0.3 & $0.86 \pm 0.06$ & $0.92 \pm 0.09$ & $0.20 \pm 0.05$ & $1.46 \pm 0.22$ & $1.45 \pm 0.13$ & $4.22 \pm 0.06$ & $0.05 \pm 0.03$\\ \hline
\end{tabular}
\end{scriptsize}
\caption{Network statistics for spontaneous LFP activity.}\label{table3}
\end{table}

\begin{table}[!h]
\centering
\begin{scriptsize}
\begin{tabular}{|c|c|c|c|c|c|c|c|c|c|c|}
\hline
\multicolumn{11}{|c|}{ pre-stimulus }\\ \hline
%\#nodes & \#edges & win (ms) & $\theta$ & $C$ & $C^{l}$ & $C^r$ & $L$ & $L^r$ & $S = \frac{C/C^r}{L/L^r}$ & $\omega = \frac{L^r}{L} - \frac{C}{C^{l}}$ \\ \hline
nodes & edges & win & $\theta$ & $C$ & $C^{l}$ & $C^r$ & $L$ & $L^r$ & $S$ & $\omega$ \\ \hline
$24$ & $199 \pm 48$ & 50 & 0.4 & $0.82 \pm 0.05$ & $0.96 \pm 0.04$ & $0.56 \pm 0.11$ & $1.49 \pm 0.11$ & $1.41 \pm 0.08$ & $1.37 \pm 0.02$ & $0.09 \pm 0.01$ \\ \hline
$24$ & $157  \pm 47$ & 50 & 0.5 & $0.89 \pm 0.06$ & $0.92 \pm 0.07$ & $0.40 \pm 0.12$ & $1.86 \pm 0.23$ & $1.63 \pm 0.01$ & $1.78 \pm 0.47$ & $-0.08 \pm 0.02$ \\ \hline
$24$ & $208 \pm 43$ & 100 & 0.3 & $0.83 \pm 0.04$ & $0.98 \pm 0.37$ & $0.62 \pm 0.08$ & $1.39 \pm 0.09$ & $1.34 \pm 0.07$ & $1.26 \pm 0.01$ & $0.11 \pm 0.00$\\ \hline
$24$ & $179 \pm 44$ & 100 & 0.4 & $0.88 \pm 0.05$ & $0.94 \pm 0.05$ & $0.47 \pm 0.10$ & $1.72 \pm 0.19$ & $1.54 \pm 0.11$ & $1.59 \pm 0.03$ & $-0.04 \pm 0.02$ \\ \hline
\multicolumn{11}{|c|}{ post-stimulus }\\ \hline
$24$ & $217 \pm 45$ & 50 & 0.4 & $0.83 \pm 0.05$ & $0.96 \pm 0.04$ & $0.55 \pm 0.10$ & $1.50 \pm 0.11$ & $1.42 \pm 0.08$ & $1.37 \pm 0.02$ & $0.07 \pm 0.01$ \\ \hline
$24$ & $156 \pm 46$ & 50 & 0.5 & $0.90 \pm 0.05$ & $0.92 \pm 0.07$ & $0.40 \pm 0.11$ & $1.89 \pm 0.26$ & $1.64 \pm 0.14$ & $1.77 \pm 0.15$ & $-0.11 \pm 0.02$ \\ \hline
$24$ & $219 \pm 40$ & 100 & 0.3 & $0.83 \pm 0.04$ & $0.98 \pm 0.03$ & $0.62 \pm 0.08$ & $1.39 \pm 0.08$ & $1.33 \pm 0.07$ & $1.26 \pm 0.02$ & $0.10 \pm 0.00$ \\ \hline
$24$ & $181 \pm 42$ & 100 & 0.4 & $0.88 \pm 0.05$ & $0.95 \pm 0.05$ & $0.46 \pm 0.10$ & $1.72 \pm 0.18$ & $1.54 \pm 0.29$ & $1.61 \pm 0.03$ & $-0.03 \pm 0.01$\\ \hline
\end{tabular}
\end{scriptsize}
\caption{Network statistics for evoked LFP activity.}\label{table4}
\end{table}

\begin{table}[!h]
\centering
\begin{scriptsize}
\begin{tabular}{|c|c|c|c|c|c|c|c|c|c|}
\hline
\multicolumn{10}{|c|}{ Computational model composed by small-world networks }\\ \hline
%\#nodes & \#edges & $\rho$ & $C$ & $C^{l}$ & $C^r$ & $L$ & $L^r$ & $S = \frac{C/C^r}{L/L^r}$ & $\omega = \frac{L^r}{L} - \frac{C}{C^{l}}$ \\ \hline
nodes & edges & $\rho$ & $C$ & $C^{l}$ & $C^r$ & $L$ & $L^r$ & $S$ & $\omega$ \\ \hline
$100$ & $1503 \pm 211$ & 0.01 & $0.76 \pm 0.06$ & $0.76 \pm 0.04$ & $0.49 \pm 0.05$ & $3.22 \pm 0.34$ & $2.71 \pm 0.31$ & $1.31 \pm 0.02$ & $-0.12 \pm 0.04$ \\ \hline
$100$ & $1588  \pm 192$ & 0.03 &  $0.87 \pm 0.02$ & $0.89 \pm 0.02$ & $0.66 \pm 0.02$ & $1.34 \pm 0.15$ & $1.33 \pm 0.15$ & $1.98 \pm 0.00$ & $0.01 \pm 0.00$ \\ \hline
$100$ & $1524 \pm 181$ & 0.05 & $0.93 \pm 0.03$ & $0.93 \pm 0.03$ & $0.70 \pm 0.03$ & $1.12 \pm 0.10$ & $1.12 \pm 0.10$ & $1.61 \pm 0.00$ & $0.00 \pm 0.00$\\ \hline
$100$ & $1395 \pm 184$ & 0.075 & $0.71 \pm 0.03$ & $0.77 \pm 0.01$ & $0.46 \pm 0.05$ & $1.49 \pm 0.06$ & $1.48 \pm 0.05$ & $3.64 \pm 0.00$ & $0.07 \pm 0.00$ \\ \hline
$100$ & $1477  \pm 156$ & 0.1 & $0.63 \pm 0.04$ & $0.72 \pm 0.04$ & $0.26 \pm 0.07$ & $1.94 \pm 0.12$ & $1.84 \pm 0.09$ & $2.29 \pm 0.12$ & $0.09 \pm 0.00$ \\ \hline
$100$ & $1501 \pm 169$ & 0.25 & $0.68 \pm 0.04$ & $0.63 \pm 0.09$ & $0.17 \pm 0.07$ & $2.33 \pm 0.17$ & $2.15 \pm 0.15$ & $3.73 \pm 0.01$ & $-0.19 \pm 0.03$\\ \hline
$100$ & $1396 \pm 205$ & 0.5 & $0.68 \pm 0.05$ & $0.62 \pm 0.07$ & $0.16 \pm 0.06$ & $2.27 \pm 0.17$ & $2.11 \pm 0.14$ & $3.81 \pm 0.03$ & $-0.21 \pm 0.04$ \\ \hline
\multicolumn{10}{|c|}{ Computational model composed by random networks }\\ \hline
$100$ & $1536 \pm 186$ & 0.01 & $0.43 \pm 0.11$ & $0.74 \pm 0.05$ & $0.44 \pm 0.10$ & $1.91 \pm 0.18$ & $1.83 \pm 0.13$ & $0.94 \pm 0.00$ & $0.36 \pm 0.01$ \\ \hline
$100$ & $1457  \pm 193$ & 0.03 & $0.39 \pm 0.11$ & $0.70 \pm 0.07$ & $0.39 \pm 0.08$ & $1.98 \pm 0.23$ & $1.87 \pm 0.17$ & $0.95 \pm 0.08$ & $0.37 \pm 0.02$ \\ \hline
$100$ & $1319 \pm 168$ & 0.05 & $0.37 \pm 0.09$ & $0.65 \pm 0.37$ & $0.32 \pm 0.08$ & $2.13 \pm 0.30$ & $1.97 \pm 0.21$ & $1.06 \pm 0.04$ & $0.34 \pm 0.02$\\ \hline
$100$ & $1209 \pm 179$ & 0.075 & $0.19 \pm 0.08$ & $0.36 \pm 0.05$ & $0.12 \pm 0.08$ & $3.11 \pm 0.64$ & $2.78 \pm 0.52$ & $1.33 \pm 0.04$ & $0.37 \pm 0.02$ \\ \hline
$100$ & $1472  \pm 162$ & 0.1 & $0.28 \pm 0.09$ & $0.54 \pm 0.07$ & $0.21 \pm 0.08$ & $2.49 \pm 0.39$ & $2.24 \pm 0.31$ & $1.17 \pm 0.06$ & $0.38 \pm 0.02$ \\ \hline
$100$ & $1466 \pm 181$ & 0.25 & $0.42 \pm 0.02$ & $0.72 \pm 0.37$ & $0.46 \pm 0.21$ & $2.16 \pm 0.33$ & $2.08 \pm 0.21$ & $0.87 \pm 0.01$ & $0.36 \pm 0.00$\\ \hline
$100$ & $1702 \pm 184$ & 0.5 & $0.38 \pm 0.07$ & $0.61 \pm 0.09$ & $0.47 \pm 0.13$ & $2.43 \pm 0.48$ & $1.94 \pm 0.19$ & $0.64 \pm 0.02$ & $-0.42 \pm 0.02$ \\ \hline\end{tabular}
\end{scriptsize}
\caption{Network statistics for the computational model. $\rho$ indicates the density (small-world networks per node) and the winsize is set to 100 ms in all analyses.}\label{table5}
\end{table}

\end{document}